# Direct soliton generation in cw-pumped doubly resonant degenerate optical parametric oscillators

**MINGMING NIE,[1,2] AND SHU-WEI HUANG[1,*]**

[1]*Department of Electrical, Computer & Energy Engineering, University of Colorado Boulder, Boulder, CO 80309, USA*
[2]*email: Mingming.Nie@colorado.edu*
*\*Corresponding author:* ShuWei.Huang@colorado.edu

**We analytically and numerically study the direct soliton generation in a cw-pumped doubly resonant degenerate optical parametric oscillator via pump frequency scanning. By means of bifurcation and linear stability analysis of the homogeneous solutions, we discriminate distinctive soliton forming mechanisms and corresponding dynamics, depending on pump and signal group velocity dispersions (GVDs). To the best of our knowledge, this is the first study regarding the dynamics of pure quadratic solitons. In addition, we discuss how to extend soliton existing regime and easily access quadratic soliton by introducing modulation instabilities through the adjustment of pump and signal GVDs. Our study will benefit the theoretical cavity design and experimental realization for pure quadratic solitons.**

Optical parametric oscillator (OPO) is intrinsically broadband and tunable and it extends the optical frequency comb (OFC) technology to otherwise inaccessible wavelengths in the mid-infrared (MIR) molecular fingerprinting spectral regime [1]. Traditionally, synchronous pumping is utilized to generate OPO-based OFC, through which viable MIR OFC sources based on periodically poled lithium niobate (PPLN) [2] and orientation-patterned gallium arsenide (OP-GaAs) [3] have been successfully implemented. However, synchronously and near-synchronously pumped OPOs require additional mode-locked lasers and associated synchronization electronics, thus generally resulting in increased complexity, large footprint, and high cost for such OPOs. To address these issues, techniques to mode-lock continuous-wave (cw) pumped OPO have been investigated and developed. Early efforts in this field focused on active mode-locking with intra-cavity modulators [4]. Recent theoretical analyses and experiments further show that OFC is attainable in a high-Q OPO [5-10], either by large walk-off induced modulation instability (MI) or by quadratic soliton mode-locking principle. Among them, OFC based on quadratic soliton is of advantage due to large bandwidth and low timing jitter, making it a competitive ultrafast laser source at the MIR spectral region.

Although the existence of pure quadratic soliton has been theoretically and numerically predicted in both degenerate OPOs (DOPOs) [7-10] and SHG cavities [11,12], so far no experiment has been demonstrated, which might be due to: (i) the relatively high demanding of the small group velocity mismatch (GVM) between the two interactive fields, which can be solved by extensive on-chip dispersion engineering; (ii) the obscure experimental route to access quadratic solitons, which is based on the theoretical understanding of system dynamics. Despite the various methods to access conventional dissipative Kerr soliton (DKS), such as frequency tuning [13, 14], power kicking [15, 16] and pulse seeding [17], no evidence shows the feasibility to apply these methods for experimental generation of quadratic soliton.

Here, we theoretically and numerically study the direct soliton generation in a cw-pumped doubly resonant DOPO (DR-DOPO) via pump frequency tuning. By means of bifurcation and linear stability analysis of the homogeneous solutions, we discriminate distinctive soliton forming mechanisms and corresponding dynamics, depending on the pump and signal group velocity dispersions (GVDs). To the best of our knowledge, this is the first study regarding the dynamics of pure quadratic solitons. In addition, we discuss how to extend soliton existing regime and easily access quadratic soliton by introducing modulation instabilities through the adjustment of pump and signal GVDs. Beyond the DR-DOPOs, the analysis can be expanded to other high-Q cavities with dominant $\chi^{(2)}$ nonlinearity. Our study will benefit the theoretical cavity design and experimental realization for pure quadratic solitons.

The field evolution in the retarded time frame through a cw-pumped DR-DOPO obeys the coupled equations [9]:

$$\frac{\partial A}{\partial z} = \left[ -\frac{\alpha_{c1}}{2} - i\frac{k_1^{''}}{2}\frac{\partial^2}{\partial \tau^2} \right] A + i\kappa BA^* e^{-i\Delta kz}, \quad \textbf{(1a)}$$

$$\frac{\partial B}{\partial z} = \left[ -\frac{\alpha_{c2}}{2} - \Delta k' \frac{\partial}{\partial \tau} - i\frac{k_2^{''}}{2}\frac{\partial^2}{\partial \tau^2} \right] B + i\kappa A^2 e^{i\Delta kz}, \quad \textbf{(1b)}$$

and the boundary conditions:

$$A_{m+1}(0,\tau) = \sqrt{1-\theta_1}\, A_m(L,\tau) e^{-i\delta_1}, \quad \textbf{(2a)}$$

$$B_{m+1}(0,\tau) = \sqrt{1-\theta_2}\, B_m(L,\tau) e^{-i\delta_2} + \sqrt{\theta_2}\, B_{in}, \quad \textbf{(2b)}$$

where $A$ is the signal field envelope, $B$ is the pump field envelope, $B_{in}$ is the cw pump, $\alpha_{c1,2}$ are the propagation losses per unit length, $\Delta k$ is the wave-vector mismatch, $\Delta k'$ is the GVM, and $k_{1,2}^{''}$ are GVD coefficients, $L$ is the monolithic cavity length, $\theta_{1,2}$ are the coupler transmission coefficients and $\delta_{1,2}$ are the pump-resonance and signal-resonance phase detuning, respectively. $\kappa$ is the normalized second-order nonlinearity coupling coefficient. Higher-order dispersion and nonlinearity are both neglected for simplicity.

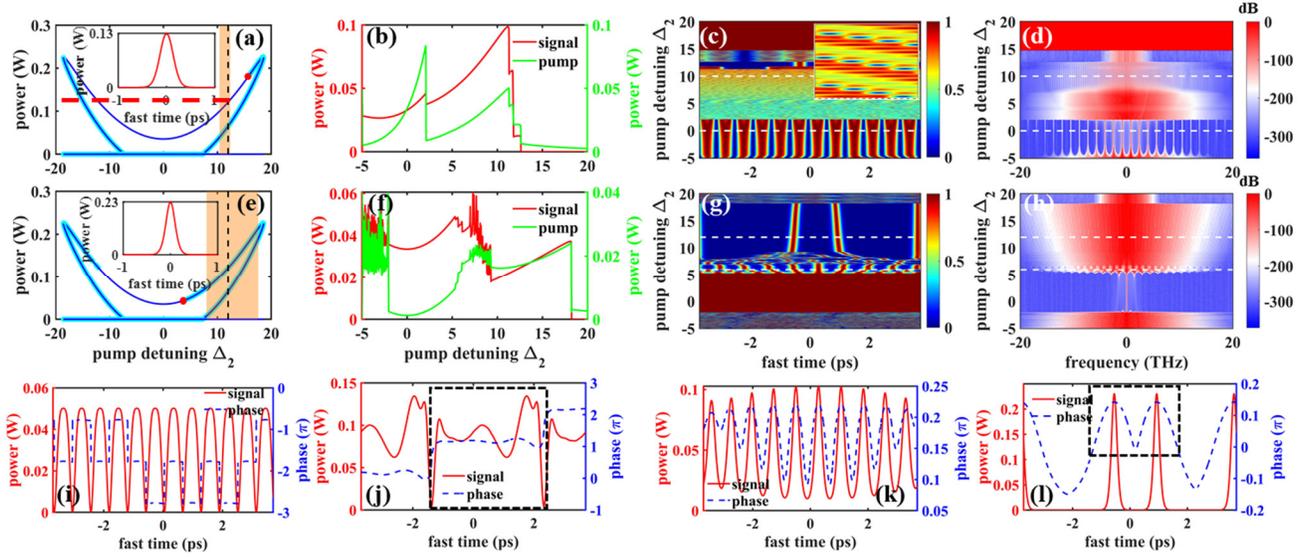

Fig. 1. Dynamics of DR-DOPO by frequency scanning with a slow speed of 70.9 GHz/ms (normalized pump detuning $\Delta_2$ from -5 to 20 within $2\times10^6$ roundtrips). $P_{in}$ =2 mW, $\alpha_1 = \alpha_2 = \theta_1 = \theta_2 = \pi/1600$, $\kappa$ = 54.2 W$^{-1/2}$m$^{-1}$, $\Delta k'$=0, $k_1^*$ =-325 fs$^2$/mm, $\xi$=0. (a)-(d), (i) and (j): pump GVD $k_2^*$ =163 fs$^2$/mm. (e)-(h), (k) and (l): pump GVD $k_2^*$ =-163 fs$^2$/mm. (a)(e) resonance diagrams and linear stability analysis for the non-zero solutions and zero solutions. The unstable solutions are indicated by cyan circles. The red points indicate where MI starts for the red pump detuning side. The black dashed lines indicate the pump detuning ($\Delta_2$=12). The light orange shaded areas indicate the soliton existing regime. The insets are the signal soliton profiles at $\Delta_2$=12. The peak powers are 0.13 W (a) and 0.23 W (e), respectively. The red dashed line in (a) indicates the cw power of 0.13 W. (b)(f) intra-cavity average power evolution, (c)(g) signal pulse evolution and (d)(h) signal spectrum evolution with $\Delta_2$. The pulse and spectrum intensities are normalized for each roundtrip to clearly show the evolution. The inset of (c) shows the zoomed-in results with $\Delta_2$ between 10 and 10.5. (i)(j) signal profiles and corresponding phase as snapshots at $\Delta_2$=0 and $\Delta_2$=10, indicating by dashed lines in (c) and (d). Dark and grey pulses are formed through domain wall locking of two stable upper branches out of phase. The dashed box in (j) shows an in-phase localized structure. (k)(l) signal profiles and corresponding phase as snapshots at $\Delta_2$=6 and $\Delta_2$=12, indicating by dashed lines in (g) and (h). The dashed boxes in (k) show in-phase localized structures.

Under the mean-field and good cavity approximations, the coupled-wave equations with boundary conditions can be simplified into a single mean-field equation for the signal field [8]:

$$t_R \frac{\partial A}{\partial t} = \left(-\alpha_1 - i\delta_1 - i\frac{k_1^* L}{2}\frac{\partial^2}{\partial \tau^2}\right)A \\ -\left(\kappa L \mathrm{sinc}(\xi)\right)^2 A^*\left[A^2 \otimes J(\tau)\right] + i\rho A^*, \quad (3)$$

where $t$ is the "slow time" that describes the envelope evolution over successive round-trips, $t_R$ is the roundtrip time, $\tau$ is the "fast time". $J(\tau) = \mathcal{F}^{-1}[\hat{J}(\Omega)]$ is the effective third-order nonlinearity and the nonlinear response function $\hat{J}(\Omega) = \left(\alpha_2 + i\delta_2 - i\Delta k' L\Omega - ik_2^* L\Omega^2/2\right)^{-1}$ describes the dispersion of the effective third-order nonlinearity [8]. Here, $\Omega$ is the offset angular frequency with respect to the signal resonance frequency. $\alpha_{1,2}$ are the total linear cavity loss for the signal and pump laser, $\xi=\Delta kL/2$ is the wave-vector mismatch parameter and $\rho$ is the phase-sensitive parametric pump driving term.

We consider a cw-pumped, group-velocity-matched ($\Delta k'$=0) and perfectly-phase-matched ($\xi$=0) DR-DOPO made of monolithic PPLN microring, with a 28-μm$^2$ mode area, a 1-mm cavity length, a 1262-nm cw pump wavelength, and a 2524-nm signal center wavelength, which can be readily achieved in the real-world parameters [9].

The bifurcation behaviors are investigated for two cases with different pump GVDs of 163 fs$^2$/mm and -163 fs$^2$/mm, as shown in Fig. 1(a) and 1(e). The resonance diagrams are obtained by finding out the homogeneous solutions (including zero and non-zero solutions) of Eq. (3), while bifurcations are obtained via linear stability analysis [5]. The non-zero solutions are given by

$$|A_0|^2 = \frac{(\delta_1\delta_2 - \alpha_1\alpha_2) \pm \sqrt{(\kappa L)^2 \theta_2 P_{in} - (\alpha_1\delta_2 + \alpha_2\delta_1)^2}}{(\kappa L)^2}, \quad (4)$$

while the homogeneous solutions exhibit MI gain for positive real part of the eigenvalues:

non-zero:
$$\lambda_\pm = -\left[\alpha_1 + \mu^2 |A_0|^2 J_+(\Omega)\right] \\ \pm \sqrt{(\alpha_1^2 + \delta_1^2) - \left[\delta_1 - D_2\Omega^2 - i\mu^2|A_0|^2 J_-(\Omega)\right]^2}, \quad (5a)$$

zero: $\lambda_\pm = -\alpha_1 \pm \sqrt{\left|\kappa L \hat{J}(0)\right|^2 \theta_2 P_{in} - (\delta_1 - D_2\Omega^2)^2}, \quad (5b)$

where $J_\pm(\Omega) = \hat{J}(\Omega) \pm \hat{J}^*(-\Omega)$, $D_2 = k_1^* L/2$, $P_{in} = |B_{in}|^2$ is the input pump power and $|A_0|^2$ is the power of non-zero solution.

The system exhibits two tilted bistable regimes for both cases [Fig. 1(a) and 1(e)], which is typical for doubly resonant quadratic cavities [6,11]. The whole lower branches and zero solution below

$\Delta_2 = 2\sqrt{\left[\alpha_1\alpha_2 - (\kappa L)^2 \theta_2 |B_{in}|^2\right]/\left[2 - (2\alpha_1 + \alpha_2)^2\right]}/\alpha_2$ are unstable for both cases, while the stabilities of the upper branches are quite different. We only consider the stability difference of the upper branch on the red-detuned side where the soliton might exist [shaded area in Fig. 1(a) and 1(e)].

For the first case of pump GVD $k_2^{'}$=163 fs²/mm, MI starts close to the end of the upper branch [red point in Fig. 1(a)]. The upper branch is modulationally stable in the soliton existing regime, which means cw solution can coexist with soliton. It is quite tricky that the soliton will not be perturbed by the cw solution. The possibility is that soliton depends on the cw solution, which indicates an alternative mechanism to form solitons: domain wall locking. Usually, domain wall is formed by connecting two stable homogeneous solutions, i. e., a stable upper branch and a stable zero solution or two stable upper branches out of phase. Once two domain walls share the same group velocity, they lock and form a localized structure. Here, the soliton generation is referred to the domain wall locking between the stable upper branch and stable zero solution, verified by the consistency between soliton peak power [inset of Fig. 1(a)] and the cw power of upper branch.

The dynamics in Figs. 1(b) to 1(d) is obtained by numerically solving Eq. (1) and Eq. (2) from noise via the standard split-step Fourier method, with pump frequency tuned from blue side ($\Delta_2 = \delta_2/\alpha_2$=-5) to red side ($\Delta_2$=20). During the tuning process, localized structures [Fig. 1(i)] start from initial noise through domain wall locking between two stable upper branches out of phase [7, 9]. According to Eq. (3), the system symmetry can be achieved by conducting the transformation $A \rightarrow -A$. Thus, two upper branches out of phase (with π difference in phase), lead to the formation of Ising walls and dark pulses (with lowest peak power of 0 W). As the detuning continues going to the red side, at around $\Delta_2$=2 the Ising walls transit to nonstationary Bloch walls with a constant drifting velocity [Fig. 1(c) inset and Fig. 2(c)] owing to the chiral symmetry breaking [18], which is denoted as Ising-Bloch transition. New localized structures and grey pulses (with lowest peak power close to 0 W but not exactly at 0 W) are formed as shown in Fig. 1(j). When the detuning goes deeply into the soliton existing regime, the grey pulses can not be kept stable and another kind of domain wall forms, which connects a stable upper branch and a stable zero solution and is usually denoted as *topological soliton*. The final soliton number are determined by the segment of the in-phase components since they will attract each other and form localized structures. According to Fig. 1(b), evident soliton step is formed during the frequency tuning process, similar to the counterpart of conventional DKS.

For the second case of pump GVD $k_2^{'}$=-163 fs²/mm, MI starts close to the resonance peak [red point in Fig. 1(e)]. When the pump frequency is tuned from the blue side to the red side, the system will experience cw, MI and then go into the soliton existing regime. The upper branch is modulationally unstable in the soliton existing regime, which is of the same picture of conventional DKS as *non-topological solitons*.

As for the dynamics in Fig. 1(f) to 1(h), cw state first starts from initial noise. Obviously, domain wall locked structures can also be obtained at different initial noise conditions [not shown in Fig. 1(g)]. When the detuning continues going to red-detuned side, Turing patterns [Fig. 1(k)] then occurs through MI from the unstable cw state or domain wall locked structures. When the detuning goes deeply into the soliton existing regime, the in-phase patterns close to each other will collapse and form solitons or disappear, while the out-of-phase patterns will repulse each other and evolve into solitons. Figure 1(l) shows three in-phase solitons among which the two solitons close to each other in the dashed box is anticipated to collapse. According to Fig. 1(f), the evident soliton step during the frequency tuning process is much longer than that in Fig. 1(b), resulting from larger soliton existing regime.

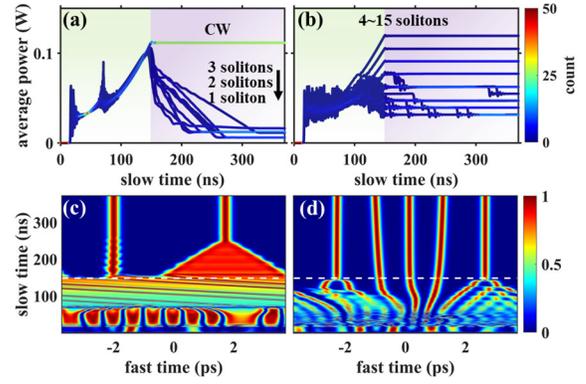

Fig. 2. (a)(c) pump GVD $k_2^{'}$=163 fs²/mm. (b)(d) pump GVD $k_2^{'}$=-163 fs²/mm. (a)(b) Histogram of 50 overlaid intra-cavity average power traces with frequency tuning at different pump GVDs. The steps are caused by soliton attraction and collision. (c)(d) Pulse evolution during the frequency tuning process. The pulse intensities are normalized for each roundtrip to clearly show the evolution. The dashed lines indicate where the detuning starts to keep at $\Delta_2$=12.

We attribute the narrower soliton existing regime for the first case to two reasons: (i) soliton perturbation resulting from the TPA peaks and phase anomalies in the upper zone where $k_1^{'} \cdot k_2^{'}$ <0 [8]. If pump GVD is much larger than 163 fs²/mm, the soliton existing regime will become narrower and eventually disappear. (ii) to sustain *topological soliton* formation, the group velocity difference of the two domain walls depends on the detuning and should be limited in a small value around the Maxwell point [6].

We also study the system dynamical behavior by repeating 50 times frequency tuning process and record the intra-cavity average power traces, as shown in Fig. 2(a) and 2(b). Each simulation starts with reinitialized noises to make sure there is no correlation between consecutive runs. The pump frequency $\Delta_2$ is linearly tuned from -5 to 12 at the first 148 ns (within 2×10⁴ roundtrips, corresponding to a fast scanning speed of 4.82 THz/ms) and then kept at $\Delta_2$=12 for soliton generation and stabilization. Note that the scanning speed is much faster than that in Fig. 1 to save the computing time. However, the fast tuning speed will not significantly affect the system dynamics and especially the final soliton state, which depends on the soliton existing regime. For the first case of pump GVD $k_2^{'}$=163 fs²/mm, it can not be guaranteed to access *topological solitons* for each scan. In fact, it has the largest probability to access the stable upper branch solutions [Fig. 2(a)]. For the other case of pump GVD $k_2^{'}$=-163 fs²/mm, the final soliton number is essentially stochastic, from 4 to 15, similar to the conventional DKS generation.

The soliton number difference for the two cases can be directly attributed to the different soliton formation mechanisms: one from domain wall locking and the other from MI. Since the first case of pump GVD $k_2^{'}$=163 fs²/mm experiences no MI, *topological soliton* generation through domain wall locking, requires at least two components out of phase in the beginning, which strongly depends on the stochastic process from noise.

In order to achieve soliton generation at each scan and avoid cw state, MI should be introduced in the soliton existing regime, i. e., changing from *topological soliton* to a *non-topological* one. Therefore, the MI starting point is important, which can be conveniently adjusted in the design phase, by controlling pump and signal GVDs through dispersion engineering. As shown in Fig. 3, to move the MI starting point close to the zero detuning, it is better to operate in the lower zone where $k_1'' \cdot k_2'' > 0$ with large pump GVD amplitude or in the upper zone where $k_1'' \cdot k_2'' < 0$ with small pump GVD amplitude. Moreover, small amplitude of signal GVD is preferred in the lower zone while large amplitude of signal GVD is preferred in the upper zone. Therefore, for broadband soliton comb generation, it is beneficial to operate in the lower zone with small signal GVD amplitude. As for the experiment, if MI regime does not overlap with the soliton existing regime, other methods apart from the frequency tuning can be introduced to successfully access the soliton, for example, power kicking, pulse seeding, mode corssings induced MI [19] and so on.

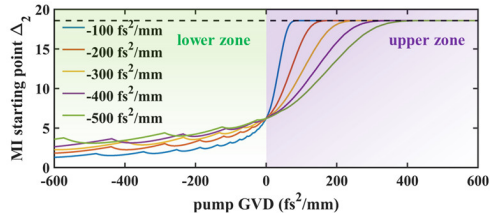

Fig. 3. MI starting point on the pump detuning axis versus pump GVD for five different signal GVDs. Note that we only care about MI starting point on the red side where signal soliton exists with anomalous signal GVD. The dashed black line indicates the detuning edge in the resonance diagram [Fig. 1(a) and Fig. 1(e)].

In conclusion, we theoretically and numerically study the direct soliton generation in a cw-pumped DR-DOPO via pump frequency tuning. By means of bifurcation and linear stability analysis of the homogeneous solutions, we discriminate distinctive soliton formation mechanisms and corresponding dynamics, depending on the pump and signal GVDs. When the upper branch of the cw solutions is modulationally stable in the soliton existing regime, quadratic solitons are formed through domain wall locking. During the pump frequency tuning process, the system might successively experience Ising walls (dark pulses), Bloch walls (grey pulses), and quadratic soliton (*topological soliton*) states, or only cw state without soliton generation, depending on the initial noise. On the other hand, when the upper branch of the cw solutions is modulationally unstable, pure quadratic soliton (*non-topological solitons*) are seeded from MI. By adjusting both the pump and signal GVDs, it is attainable to introduce MI in the soliton existing regime and easily access broadband quadratic solitons. Beyond the DRDOPOs, the analysis can also be extended to other high-Q cavities with dominant $\chi^{(2)}$ nonlinearity. Our study will benefit the theoretical cavity design and experimental realization for pure quadratic solitons.

**Disclosures**. The authors declare no conflicts of interest.